\documentclass[conference]{IEEEtran}
\IEEEoverridecommandlockouts
\usepackage{cite}
\usepackage{amsmath,amssymb,amsfonts}
\usepackage{algorithmic}
\usepackage{graphicx}
\usepackage{textcomp}
\usepackage{xcolor}
\usepackage{soul}
\usepackage{booktabs} 
\usepackage{multirow}
\usepackage{array}
\def\BibTeX{{\rm B\kern-.05em{\sc i\kern-.025em b}\kern-.08em
    T\kern-.1667em\lower.7ex\hbox{E}\kern-.125emX}}
\makeatletter
\newcommand{\linebreakand}{%
  \end{@IEEEauthorhalign}
  \hfill\mbox{}\par
  \mbox{}\hfill\begin{@IEEEauthorhalign}
}
\makeatother
\begin{document}

\title{A Pathway to Near Tissue Computing through Processing-in-CTIA Pixels for Biomedical Applications\\
% {\footnotesize \textsuperscript{*}Note: Sub-titles are not captured for https://ieeexplore.ieee.org  and

}

\author{
\IEEEauthorblockN{1\textsuperscript{st} Zihan Yin}
\IEEEauthorblockA{\textit{ECE} \\
\textit{University of Wisconsin-Madison}\\ 
Madison, USA\\
zyin83@wisc.edu}
\and
\IEEEauthorblockN{2\textsuperscript{nd} Subhradip Chakraborty}
\IEEEauthorblockA{\textit{ECE} \\
\textit{University of Wisconsin-Madison}\\
Madison, USA \\
chakrabort42@wisc.edu}
\and
\IEEEauthorblockN{3\textsuperscript{rd} Ankur Singh}
\IEEEauthorblockA{\textit{ECE} \\
\textit{University of Wisconsin-Madison}\\
Madison, USA \\
ankur.singh@wisc.edu}
\linebreakand
\IEEEauthorblockN{4\textsuperscript{th} Chengwei Zhou}
\IEEEauthorblockA{\textit{ECSE} \\
\textit{Case Western Reserve University}\\
Cleveland, USA\\
chengwei.zhou@case.edu}
\and
\IEEEauthorblockN{5\textsuperscript{th} Gourav Datta}
\IEEEauthorblockA{\textit{ECSE} \\
\textit{Case Western Reserve University}\\
Cleveland, USA\\
gourav.datta@case.edu}
\and
\IEEEauthorblockN{6\textsuperscript{th} Akhilesh Jaiswal} %Given Name Surname}
\IEEEauthorblockA{\textit{ECE} \\
\textit{University of Wisconsin-Madison}\\ 
Madison, USA\\
akhilesh.jaiswal@wisc.edu}
}

\maketitle

\begin{abstract}
% 1. Summarize the current camera functionality and mention the benefit on in-pixel computing, near tissue computing\\
% 2. Introducing the CTIA in-pixel computing design\\
% 3. Highlight the result and performance improvement on linearity, data set evaluation result.
Near-tissue computing requires sensor-level processing of high-resolution images, essential for real-time biomedical diagnostics and surgical guidance. To address this need, we introduce a novel Capacitive Transimpedance Amplifier-based In-Pixel Computing (CTIA-IPC) architecture. Our design leverages CTIA pixels that are widely used for biomedical imaging owing to the inherent advantages of excellent linearity, low noise, and robust operation under low-light conditions. We augment CTIA pixels with IPC to enable precise deep learning computations including multi-channel, multi-bit convolution operations along with integrated batch normalization (BN) and Rectified Linear Unit (ReLU) functionalities in the peripheral ADC (Analog to Digital Converters). This design improves the linearity of Multiply and Accumulate (MAC) operations while enhancing computational efficiency. Leveraging 3D integration to embed pixel circuitry and weight storage, CTIA-IPC maintains pixel density comparable to standard CTIA designs. Moreover, our algorithm-circuit co-design approach enables efficient real-time diagnostics and AI-driven medical analysis. Evaluated on the EndoVis tissue dataset (1280$\times$1024), CTIA-IPC achieves approximately $\mathbf{12\times}$ reduction in data bandwidth, yielding segmentation IoUs of 75.91\% (parts), and 28.58\% (instrument)—a minimal accuracy reduction ($\sim$1.3\%–2.5\%) compared to baseline methods. Achieving 1.98 GOPS throughput and 3.39 GOPS/W efficiency, our CTIA-IPC architecture offers a promising computational framework tailored specifically for biomedical near-tissue computing.
% Furthermore, the architecture delivers a throughput of 1.42 GOPS and an energy efficiency of 3.56 GOPS/W. These results demonstrate that the CTIA-IPC architecture offers a promising high-performance computational framework for widely used CTIA cameras tailored specifically for biomedical near-tissue computing applications. 
% By performing operations directly within the pixel array, CTIA-IPC minimizes data transfer overhead and power consumption, making it suitable for biomedical imaging applications.
% By providing enhanced linearity, lower noise, and accurate in-situ image processing capabilities, CTIA-IPC facilitates faster, more precise tissue analysis while substantially alleviating bandwidth limitations inherent to traditional sensor architectures.

% At the core of CTIA-IPC is a bitcell design that incorporates a 4-bit SRAM based in-memory context addressable memory (CAM), which translates stored weights into the time domain, allowing direct voltage sampling from the CTIA pixel for computation.
\end{abstract}

\begin{IEEEkeywords}
CTIA, In-Pixel Computing, Near-Tissue Imaging, 3D Integration
\end{IEEEkeywords}

\section{Introduction}
% 1. Discuss trends in computational imaging and the push for edge computing in image sensors.\\
% 2. Emphasize the need for reduced data transfer and low-power, high-speed processing in medical use scenarios in near tissue imaging.\\
% 3. Overview of traditional imaging pipelines (3T,4T pixel design vs. in-pixel computing).
% Brief mention on relevant literature (P2M)cand its disadvantages (e.g., non-linearity).\\
% 4. Introduce the concept of the CTIA pixel\\
% 5. Present the novel in-pixel computing scheme that utilizes the CTIA\\
% 6. Outline the main advantages(low noise, high linearity, power)\\
% 7. Organization of the paper

\begin{figure*}[!h]
\includegraphics[width=1\linewidth]{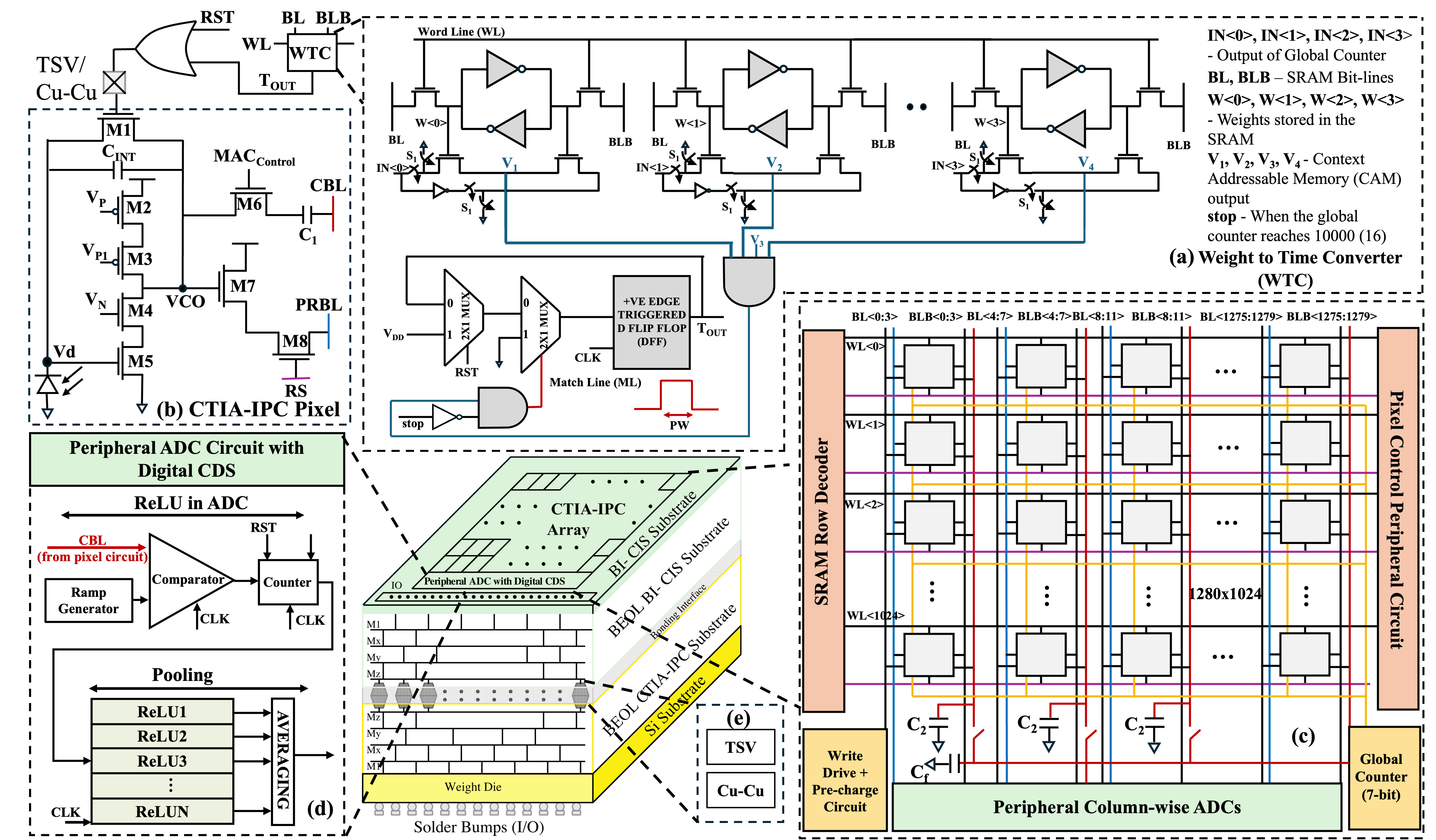}
\caption{ CTIA-IPC architecture where (a) 4-bit Weight to time converter (WTC); (b) CTIA-IPC pixel unit circuit diagram for both multiplication and regular readout; (c) 1280$\times$1024 CTIA-IPC Array with peripheral readout circuits; (d) Block diagram of Single Slope Analog to Digital Converter (SSADC) with Digital CDS architecture for ReLU and Pooling functionalities; (e) is the connection between the two dies using either Through-Silicon Vias
(TSV) or Copper-Copper bonding (Cu-Cu) for 3D integration.}
\label{fig:CTIA_Circuit}
\vspace{-8mm}
\end{figure*}
\begin{figure}[!h]
\includegraphics[width=\linewidth]{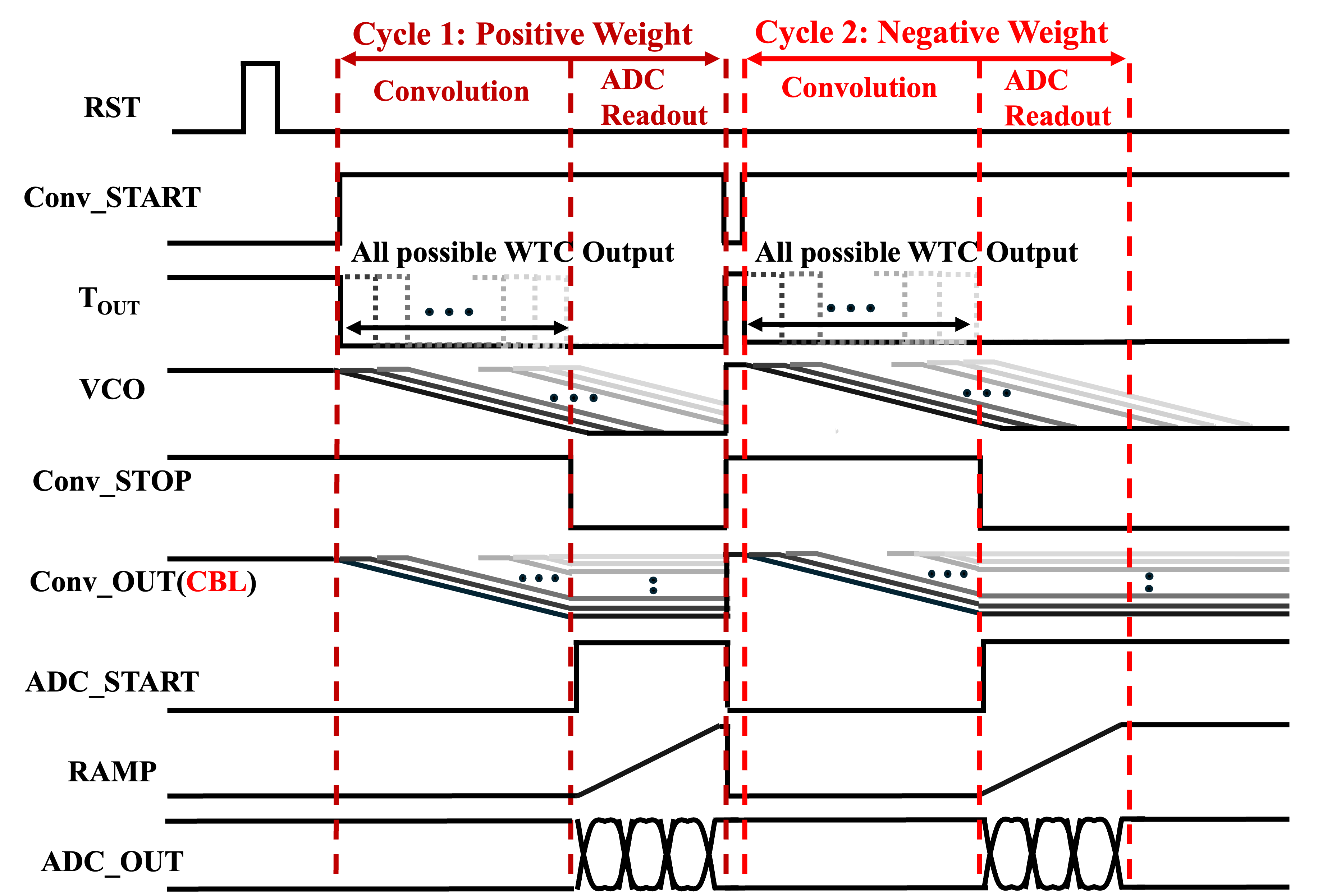}
\caption{A typical timing diagram for CTIA-IPC's convolution computation showing double sampling (one for positive and one for negative weights) including Global Counter, CTIA-IPC unit and ADC's control signals: $RST$, $Conv\_START$, $Conv\_STOP$, $ADC\_START$ and the output signals from these blocks: $T_{OUT}$, $VCO$, $Conv\_OUT$, $RAMP$ and $ADC\_OUT$.}
\label{fig:CTIA_signal}
\vspace{-5mm}
\end{figure}

The field of computational imaging is rapidly evolving, driven by a significant push towards integrating advanced edge computing capabilities directly within image sensors. This paradigm shift aims to address the growing challenges related to data transmission bottlenecks, power consumption constraints, and the demand for real-time processing speeds. In particular, embedding computation within image sensors is critical in biomedical and near-tissue computing applications, where the immediacy and precision of data processing directly correlate with patient outcomes. Furthermore, biomedical applications such as endoscopic surgery, which involves inserting a
flexible tube with a camera and light (endoscope) into the
body to visualize and operate on internal structures \cite{MayoClinicMIS,ClevelandClinicMIS}, pose significant resource, power, and bandwidth constraints for endoscopic cameras.
Traditional imaging systems, which rely heavily on off-chip data transfer and external processing, often suffer from latency, inefficiencies, and increased power overhead, significantly limiting their effectiveness in real-time medical imaging scenarios. As a result, researchers and industry professionals have progressively moved towards developing sensor-embedded computational solutions, ensuring efficient, timely, and accurate data interpretation close to the sensor.

One notable approach explored in recent computational imaging literature for computing-in-sensor is the Processing-in-Pixel (PIP) methodology\cite{aps_p2m,9785835,scamp2020eccv,roohi2023pipsim,chen2020pns,datta2022towards,yin2023design}. These aim to reduce the latency and energy overhead for convolution operations by incorporating memory and computational units within individual pixels, thereby performing early-stage computation in-situ. However, these approaches has notable limitations, \cite{aps_p2m,datta2022towards,yin2023design} has nonlinearity in output analog voltage which affects the scalability and bit-precision of the pixel circuit. The architectures described in \cite{scamp2020eccv,roohi2023pipsim,chen2020pns} exhibit poor performance under low-light conditions, significantly limiting their suitability for biomedical imaging applications. Such inherent challenges limit the practical use of the designs, particularly for high-stakes medical imaging applications where linearity, and accuracy are paramount. The constraints associated with current in-pixel processing technologies therefore require exploring more robust and accurate pixel-level computational architectures. 

To address these challenges, we introduce a Capacitive Transimpedance Amplifier-based In-Pixel Computing (CTIA-IPC) architecture, leveraging the well-established CTIA pixel's inherent linearity, low-noise characteristics, and high sensitivity under low-light conditions~\cite{zhang202320mum,5738700,zou2024low}. Building upon the advantageous properties of the CTIA pixel, our novel computational scheme CTIA-IPC integrates convolution capabilities for neural networks directly within each pixel, reducing the reliance on external data transfers and off-chip computation.

% In response to these challenges, this paper introduces an innovative in-pixel computational approach leveraging Capacitive Transimpedance Amplifier pixels (CTIA-IPC), which have traditionally been utilized for their superior linear response characteristics and enhanced sensitivity for low-light conditions\cite{zhang202320mum,5738700,zou2024low}. The CTIA pixel architecture, fundamentally designed around a charge-integration amplifier configuration, inherently offers superior linearity, dynamic range, and noise performance compared to conventional in-pixel computation methodologies. Building upon the advantageous properties of the CTIA pixel, our novel computational scheme CTIA-IPC integrates convolution capabilities for neural network use directly within each pixel, effectively reducing the reliance on external data transfers and off-chip computation.

To the best of our knowledge, this work is the first to present a CTIA-based IPC accelerator suitable for biomedical applications. The key contributions of this paper are summarized as follows:

\begin{itemize}
    \item We propose the CTIA-IPC architecture, which supports both high-resolution conventional pixel readout operations and integrated in-pixel convolution processing. 
    
    \item We implement batch normalization (BN) and Rectified Linear Unit (ReLU) activation functionality within the peripheral ADC circuit, enabling direct on-chip Convolution Neural Network (CNN) processing essential for practical biomedical imaging scenarios. 
    
    % Specifically, this design facilitates real-time feature extraction and classification tasks, critical to medical imaging applications such as on-chip tissue segmentation, anomaly detection, and immediate image-based diagnostics.
    
    \item We introduce a novel weight-to-time converter (WTC) that maps neural network weights to different exposure times, allowing the pixel to perform multiplication of input and weight for convolution efficiently.

    \item Applying 3D integration techniques, neural network weights are densely stored within vertically integrated SRAM cells through TSV/Cu-Cu bonding, allowing the CTIA-IPC architecture to maintain pixel-area density comparable to conventional CTIA pixel designs without incurring additional area overhead.  

    \item By embedding robust convolution operations directly at the pixel level, our design achieves a $12\times$ reduction in bandwidth, a computational throughput of 1.98 GOPS, and an energy efficiency of 3.39 GOPS/W, while maintaining comparable power consumption to conventional CTIA-based pixel designs that lack computation capabilities.
    
    % \item Furthermore, by minimizing external processing dependencies and reducing associated latencies, our approach is particularly suited for real-time biomedical imaging applications. These include implantable sensors, wearable diagnostic devices, and near-tissue computational modules, where rapid and precise image processing is crucial for improving medical outcomes.
    % \textcolor{red}{this contribution needs to add algorithm side}. 
    \item Furthermore, our algorithm-hardware co-design framework, incorporating 4-bit quantization and linearity-aware training, ensures robust segmentation performance under analog-domain constraints. It achieves competitive accuracy on the EndoVis dataset, with a minor performance drop of $\sim$1.3\%–2.5\% compared to baselines.  
% \vspace{-5mm}
\end{itemize}
For this paper, Section \ref{sec:CTIAarch} discusses the CTIA-IPC architecture in detail, Section \ref{sec:DataImp} talks about the medical dataset we tested with the hardware, Section \ref{sec:result} elaborates on the simulation results of the purposed circuit and Section \ref{sec:conclusion} concludes and provides some key discussions of the paper.

\section{Proposed CTIA Pixel Architecture for In-Pixel Computing}\label{sec:CTIAarch}
% 1. Design Concept: Present the whole architecture and describe its overall MAC functionality.\\
% 2. Circuit-Level Details: Describe in more detail for each part of the design (e.g., how the weight is implemented). Use timing waveform, and schematic figures here.\\
% 3. Computation flow: Walk through a signal processing cycle: from photo-charge integration to analog/digital conversion.

Commercial smartphone cameras are based on 3 or 4 transistor active pixel sensors \cite{scott2023trends,murakami20224}, while these pixels provide high fill factor they perform poorly in low light conditions and exhibit poor linearity. As such, these pixels are unsuitable for scientific applications. Owing to their high linearity and excellent low-light performance, CTIA pixel-based cameras are widely used for scientific applications including biomedical applications \cite{jiang2018nano,tran2024cmos}. In the following sub-sections we present augmenting CTIA pixels with in-situ compute capabilities targeted towards biomedical deep learning applications.

% This section describes the proposed weight to time converter (WTC) and the in-pixel CTIA architecture and its design parameters. In a CTIA-IPC unit, 6T SRAM cells are used to store weights and  pass transistor-based logic is used to perform the multiplication operation between input activation photodiode current (I) and weights (W) and a capacitive bit-line (CBL) is used to accumulate the voltage. Thanks to the active CTIA amplifier per pixel the convolution operation exhibits excellent linearity.
% We propose a novel CTIA-IPC architecture that embeds Multiply-Accumulate (MAC) capabilities directly within individual pixels, our design leverages the inherent advantages of Capacitive Transimpedance Amplifier (CTIA) pixels, particularly their superior linearity and low-noise performance. The architecture incorporates embedded SRAM cells to store convolutional kernel weights and employs pass-transistor-based logic for in-situ computation. Below, we first detail the pixel-level circuit implementation, including the CTIA pixel design, introduce our novel Weight-to-Time Converter (WTC), and subsequently outline the computational flow from charge integration to final analog-to-digital conversion.
\subsection{Proposed CTIA-IPC Design}\label{sec:CTIA_circuit}
\subsubsection{CTIA-IPC Pixel Circuit Design}
Fig. \ref{fig:CTIA_Circuit}(b) illustrates the CTIA pixel circuit, wherein the photodiode is held at a constant bias voltage. The circuit employs an operational transconductance amplifier (OTA)—implemented as a single-ended cascade common-source amplifier—with a gain of approximately 80 dB and a gain-bandwidth product (GBW) of 100 MHz. The OTA’s bias voltages are generated off-chip and distributed uniformly across the array. During the reset phase, the OTA output node $VCO$ is set to a reset voltage ($V_{RST}$) which corresponds to the bias voltage of the OTA input transistor. Upon deactivation of the M1 nMOS reset switch, a negative charge is injected into the Vd integration node, resulting in an immediate positive jump in $VCO$. Importantly, the diode current is defined as negative because the subsequent negative slope of $VCO$ during integration indicates a net discharge of the integration capacitor. At the end of the integration period, node $VCO$ holds the integration charge which could go through either CBL for MAC purpose or through PRBL for regular CTIA pixel readout. This is achieved by pulling up either line $MAC_{CONTROL}$(for MAC) or line $RS$(for regular readout).
% the switch at the end of CBL line along each column is closed at the same time which stores the integration charge in the capacitor $C_f$ as shown in Fig. \ref{fig:CTIA_Circuit}(c).

\subsubsection{Weight-Time Converter (WTC)}\label{sec:WTC}

Neural network weights in our design are stored within SRAM cells that are vertically integrated using 3D integration techniques as shown in Fig. \ref{fig:CTIA_Circuit}(e), enabling dense storage without incurring additional pixel area overhead. To facilitate IPC, these SRAM cells are reconfigured as content-addressable memory (CAM) units \cite{c2024area} capable of performing rapid matching operations. Specifically, the WTC, depicted in Fig. \ref{fig:CTIA_Circuit}(a), utilizes four CAM blocks that compare stored 4-bit SRAM weights against a multi-bit global counter value. The CAM cells adopt an 8T SRAM-based architecture to enable efficient in-memory XNOR operations. During write operations, switch $S_1$ is activated, allowing standard weight storage into conventional 6T SRAM cells. For computation, $S_1$ is deactivated to permit bitwise matching with the global counter outputs. Match signals from individual CAM blocks are combined using AND logic gate, producing a final match line that activates only when all bits align. This matched signal subsequently triggers sequential logic circuits to generate a pulse whose width accurately encodes the stored 4-bit weight. Multiplexers integrated within the sequential logic ensure proper handling of reset signals.
The resolution or step size of the multi-bit global counter determines the time at which the match line goes high and in turn controls the pulse width of the time-varying signal, thereby controlling the overall precision of the WTC.
% The output pulse width of the WTC is determined by the step size of the 7-bit global counter, specifically by selecting which four bits to use. This selection directly controls the match line (ML), thereby controlling the timing characteristics.

% The Weight-to-Time Converter (WTC) comprises of four content-addressable memory (CAM) blocks that perform matching operations between a multi-bit global counter and weights stored in SRAM, as illustrated in Fig. \ref{fig:CTIA_Circuit}(a). The CAM circuit utilizes an 8T SRAM architecture to enable in-memory XNOR operations. In this design, conventional 6T SRAM cells are employed to store 4-bit weights, while two additional transistors facilitate the matching process. During the write operation the switch S1 is switched on and during the XNOR operation S1 is switched off to allow the output bit of the global counter to match. Once the matching results are obtained from each CAM unit, they are propagated through a series of AND logic gates to compute the final match, ensuring that all bits align. The matched line is subsequently processed through a sequential logic to generate a pulse width corresponding to the 4-bit weights stored in the SRAM cells. Additionally, multiplexers are incorporated within the sequential logic to manage the reset signal. The resolution or step size of the multi-bit global counter determines the time at which the match line goes high and in turn controls the pulse width of the time-varying signal, thereby controlling the overall precision of the WTC.

\subsection{CTIA In-Pixel Compute Accelerator}

The CTIA in-pixel accelerator consists of a 1280$\times$1024 CTIA-IPC pixel array (Fig. \ref{fig:CTIA_Circuit}(c)), with each pixel comprising a standard CTIA pixel integrated with a WTC and a capacitive charge accumulation bitline (CBL), as shown in Fig. \ref{fig:CTIA_Circuit}(b). Each pixel operates in two distinct modes: (1) conventional high dynamic range (HDR) imaging, utilizing a standard 3T source-follower readout, and (2) an in-pixel MAC computation mode. In MAC mode, the WTC generates a 4-bit weighted timing signal ($T_{OUT}$), triggering the CTIA pixel's integration node ($VCO$) reset at different timestamps based on the stored weight. As illustrated in Fig. \ref{fig:CTIA_signal}, signals such as $Conv\_START$ initiate a global counter for controlling weight exposure times, while $RST$ resets both the counter's value and the pixel's integration node $VCO$'s voltage. Unlike conventional CTIA operation, where the photodiode current results in an increase in voltage, the current in this architecture is reversed. Causing the integration voltage at node $VCO$ to decrease from $V_{RST}$ toward 0V, rather than increasing toward $V_{DD}$. This design choice enhances voltage headroom and signal margin between successive MAC operations, thereby improving computational accuracy during IPC.

CNN layers typically include both positive and negative weights \cite{s2025analysis}; however, the WTC method described earlier (Section \ref{sec:WTC}) only supports positive weights. To overcome this, we repurpose the on-chip digital correlated double-sampling (CDS) circuit commonly found in commercial CMOS Image Sensors \cite{cho2012low}. Digital CDS is typically implemented with column-parallel single-slope ADCs (SS-ADCs), consisting of a ramp generator, comparator, and counter (Fig. \ref{fig:CTIA_Circuit}(d)). In standard image sensor applications, CDS measures pixel reset noise and subtracts it by first counting upwards, then downwards between two samples. Leveraging this inherent up/down counting capability, our CTIA-IPC architecture encodes positive weights through upward counting and negative weights through downward counting, accurately capturing both positive and negative contributions during pixel-level convolution.
% As weights in a CNN layer span positive and negative values, and the aforementioned WTC (Section \ref{sec:WTC}) can only convert positive weight values. We address this limitation by repurposing the on-chip digital correlated double-sampling (CDS) circuit present in many state-of-the-art commercial CMOS Image Sensors\cite{cho2012low}. Typically, a digital CDS is implemented alongside column-parallel single-slope ADCs (SS-ADCs). A single-slope ADC comprises a ramp generator, a comparator, and a counter. The comparator compares the input analog voltage with the ramp voltage as shown in Fig. \ref{fig:CTIA_signal}, which increases at a fixed rate. The counter, initially reset, increments on each clock cycle until the ramp voltage surpasses the input voltage. At that moment, the counter’s value is latched, producing the digitized value of the input analog voltage. In traditional CIS applications, the digital CDS takes two correlated samples at two different time points. The first sample captures the pixel’s reset noise, and the second sample records the actual signal plus this noise. By subtracting these two samples, the CDS removes the reset noise. In the case of an SS-ADC, this subtraction is achieved by counting up during the first sample and counting down during the second. Leveraging this built-in up/down capability in the digital CDS SS-ADC allows CTIA-IPC to measure both positive and negative weight contributions accurately as the counting up of the counter will be used for positive weights and the counting down of the counter will be used for negative weights.
Therefore every MAC operation using CTIA-IPC requires two cycles to account for positive and negative weights. For one cycle of the CTIA-IPC MAC operation, it has two phases.
\begin{enumerate}
    \item{Write Phase}: In the write phase, the SRAMs inside the CTIA-IPC unit are being written using a regular SRAM write-in method. At the end of the write phase, the neural network kernel weights are stored in dedicated SRAM cells which are connected to each pixel by Cu-Cu hybrid bonding.
    \item{Computation Phase}: In the computation phase, all pixel columns operate in parallel. The multiplication between the stored weights using WTC and the input activation, represented by the current through the photodiode, is performed across all pixels in a column. The resulting charge accumulates on the CBL, which is shared across all pixels in a given column. At the end of each CBL, a dedicated capacitor stores the final accumulated voltage. Metal-oxide-metal (MOM) capacitors are employed due to their robustness against process variations and their ability to be integrated on top of the existing CTIA architecture, thereby minimizing area overhead. Following charge accumulation, the stored voltage is processed through a switching matrix that enables charge-domain summation across multiple columns. The configuration of the switching matrix is determined by the kernel size of the neural network being implemented. The aggregated voltage is then fed into a 6-bit CDS SS-ADC. After the SS-ADC digitizes the CBL voltage through two consecutive cycles corresponding to positive and negative weights, the quantized ReLU activation is inherently realized within the SS-ADC by clipping negative digital outputs to zero. If required by the specific algorithm, the data can then pass through an on-chip pooling layer, as illustrated in Fig. \ref{fig:CTIA_Circuit}(d). This fully processed output is then ready to be forwarded to the subsequent neural network layers for further computation.
\end{enumerate} 
%Interestingly, by repurposing the on-chip CDS mechanism, our design can effectively encode both positive and negative weights, enabling the straightforward integration of ReLU within the SS-ADC.
The ReLU operation inherently sets negative outputs to zero, aligning well with biomedical imaging applications that require precise differentiation between relevant signals and background noise. Additionally, by initializing the CDS stage with values corresponding to the scale parameters from the BN layers commonly found in convolutional neural networks, our approach facilitates direct incorporation of BN functionality. Consequently, our novel CTIA-IPC architecture efficiently integrates multi-pixel convolution computations, BN, and activation functions entirely within the pixel-level ADC circuits, effectively addressing the computational demands of critical medical imaging tasks such as rapid anomaly detection, lesion segmentation, and real-time image classification.

%\subsubsection{Parallel CTIA Convolution Readout}\label{sec:switchmatrix}

To achieve parallel in-situ multi-pixel convolution, for one cycle the pixel array will activate the total number of $\frac{(i-k+2\times p)}{s\times lcm(k,s)}\times k^2\times 4$ pixels simultaneously, where $i$ denotes the spatial dimension of the input image, $k$, $p$, $s$ denote the kernel size, padding and stride of the in-pixel convolutional layer, and $\frac{(i-k+2\times p)}{s\times lcm(k,s)}$ corresponds to the maximum number of output nodes that can be calculated at the same time, $k^2 \times 4$ corresponds the filter size. (4 because of RGGB channels). For each activated pixel, the output is modulated by the photodiode current and the SRAM. The output for each column is accumulated across the CBL followed. The CBL of these columns are then accumulated through a capacitive switching logic. %which controls the accumulation across various columns before feeding the result into an ADC for binarization. The switching logic operation depends on the number of kernels and the stride size of the convolutional neural network (CNN).
A capacitive accumulation technique is employed, following the equations:
\begin{equation}
    V_{\text{ADC\_IN}} = \frac{V_1 + V_2 + \dots + V_N}{4 + \frac{2C_2}{C_1} + \frac{C_F}{C_1}}
\end{equation}

Where $V_1$, $V_2$, to $V_N$ are the multiplicative values generated by each CTIA-IPC unit. 

\begin{figure*}[!h]
\includegraphics[width=1\linewidth]{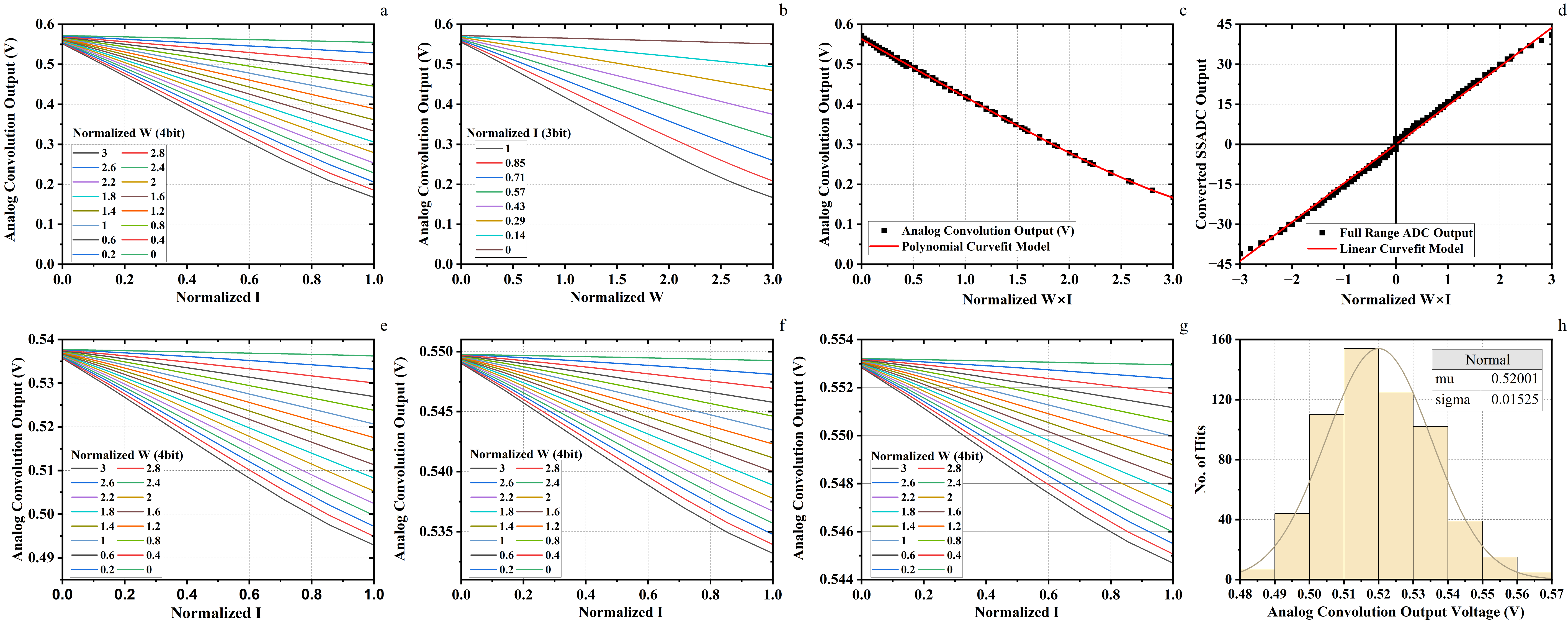}
\caption{CTIA circuit simulation results where (a) and (b) are the analog convolution output v.s. normalized weight W and input current I (representing light intensity), in (a) each line corresponds to different digital weight values that are translated from the WTC in the unit CTIA-IPC and in (b) each line corresponds to a specific value of input current. Scatter plots (c) and (d) shows linear curvefit function of the Convolution output value v.s. normalized W $\times$ I, where in (c) the convolution output data is analog voltage from line CBL shown in Fig. \ref{fig:CTIA_Circuit}(b) and in (d) is the converted digital SSADC output value. (e), (f) and (g) are the analog convolution output data v.s. sweeping one unit IPC's input I and W, where in (e) there are total of $3\times 3$ unit pixels, in (f) $5\times 5$ units and in (g) $7\times 7$ units. And (h) is the bar plot of monte carlo simulation result of the analog convolution output voltage.}
\label{fig:CTIA_simresults}
% \vspace{-5mm}
\end{figure*}

\section{Medical Imaging Application}\label{sec:DataImp}

\subsection{Details of the Dataset \& Tasks}

Minimally invasive surgery (MIS) encompasses surgical techniques that limit the size of incisions, thereby reducing wound healing time, associated pain, and risk of infection. Endoscopic surgery, a subset of MIS, involves inserting a flexible tube with a camera and light (endoscope) into the body to visualize and operate on internal structures through small incisions \cite{MayoClinicMIS,ClevelandClinicMIS}. To advance the capabilities of MIS, particularly in robotic-assisted procedures, this study focuses on surgical instrument segmentation using CNN-based models trained on a popular robot-surgery-segmentation dataset, termed \textit{Endovis} \cite{Allan2017RoboticIS}. This dataset, obtained from implantable image sensors, is well-suited for near-tissue computing, where power efficiency is crucial due to the stringent power constraints of implantable devices. In-pixel computation enabled by the CTIA-based approach is ideal for such low-power environments, as it allows feature extraction directly in the pixels, minimizing the computational load. Furthermore, the dataset’s high-contrast regions and illumination artifacts demand stable early-stage feature extraction, which the linearity provided by CTIA-based in-pixel computing ensures. This preserves critical spatial details, enhancing segmentation accuracy and robustness to lighting variations and occlusions.

The dataset consists of $8{\times}225$ high-resolution endoscopic images ($1280{\times}1024$) with pixel-wise annotations for surgical instruments. The dataset is valuable for evaluating segmentation performance in robotic-assisted surgery, with three progressively challenging subtasks: binary segmentation, parts segmentation, and instrument segmentation. These tasks involve segmenting surgical tools from complex backgrounds with occlusions, specular reflections, and fine structural details.

\subsection{Algorithm-Hardware Co-Design}

The CTIA-IPC circuit, described in Section II, implements analog convolution using non-linear transistors in modified memory-embedded pixels. Unlike prior works that ignore transistor non-idealities \cite{rxnn,memtorch}, we simulated the CTIA-IPC circuit with 22nm GlobalFoundries transistors, varying parameters like transistor width and photodiode current. The resulting SPICE outputs were modeled with a curve-fitting function as shown in Fig. \ref{fig:CTIA_simresults}(c), which replaces the convolution operation in the first network layer, similar to \cite{aps_p2m,aps_p2m_detrack}. We accumulate the function's output for each pixel in the receptive field (147 for a $7{\times}7$ kernel with 3 input channels) to model the in-pixel convolution inner-product. To compensate for analog variations, we followed this custom convolution layer with a BN layer, then applied a ReLU operation. The ReLU output is quantized to 4 bits to reduce CTIA pixel bandwidth and is transmitted off-chip for processing the remaining layers. The BN layer is fused with the preceding convolution and succeeding ReLU layers. Consider a BN layer with trainable parameters $\gamma$, $\beta$, and running mean and variance $\mu$, $\sigma$. This BN layer implements a linear function during inference:
\begin{equation}
Y = \gamma \frac{X - \mu}{\sqrt{\sigma^2 + \epsilon}} + \beta = \left(\frac{\gamma}{\sqrt{\sigma^2 + \epsilon}}\right) X + \left(\beta - \frac{\gamma \mu}{\sqrt{\sigma^2 + \epsilon}}\right)
\end{equation}
We fuse the scale term $A = \frac{\gamma}{\sqrt{\sigma^2 + \epsilon}}$ into the convolutional layer weights, adjusting the pixel weight tensor to $A \cdot \theta$, where $\theta$ is the trained weight tensor. Additionally, we shift the analog comparator trip point by $B = \left(\beta - \frac{\gamma \mu}{\sqrt{\sigma^2 + \epsilon}}\right)$, so the ReLU activation threshold becomes $-B$. This framework was used to optimize CNN training on the Endovis dataset.

Our algorithm builds upon TernausNet \cite{DBLP:journals/corr/abs-1801-05746} through hardware-algorithm co-design, replacing the VGG11 \cite{simonyan2015deepconvolutionalnetworkslargescale} encoder's initial convolution layer (with a $3 \times 3$ kernel, stride 1, and 64 channels) with the custom CTIA convolution module ($7 \times 7$ kernel, stride 2, and 16 channels), and adding a batch normalization layer as described. By omitting the subsequent max-pooling layer used in TernausNet \cite{DBLP:journals/corr/abs-1801-05746}, we preserve fine structural downsampling details. We train our custom Ternausnet network using the Adam\cite{kingma2017adammethodstochasticoptimization} optimizer with an initial learning rate of \(1e{-}4\), trained for $20$ epochs on center-cropped images. The learning rate is decayed to \(1e{-}5\) after $10$ epochs. For binary segmentation, all instrument classes are merged into a single foreground mask, optimized using Jaccard loss. For parts segmentation and instrument segmentation, a weighted cross-entropy loss is applied additionally.

\section{Results and Discussion}\label{sec:result}
% 1. Here include all the error plot(curvefit for the circuit), accuracy plot(dataset), W v.s. VOUT, I v.s. VOUT, W*I v.s. VOUT, power, \\
% 2. Analyze the data, talk about what each plot shows.

\begin{table*}[!ht]
\begin{center}
\scriptsize\addtolength{\tabcolsep}{-0pt}
\resizebox{\linewidth}{!}{ 
\begin{tabular}{lccccccccc}
\toprule
\multicolumn{1}{r}{System Overview}&\multicolumn{2}{c}{Technical Details} &\multicolumn{5}{c}{CNN Support} &\multicolumn{2}{c}{Performance Metrics} \\
\cmidrule(lr){2-3} \cmidrule(lr){4-8} \cmidrule(lr){9-10}
Works &Supported Func. &Tech Node &Weight Form & Reconfigurability & BN& ReLU&Conv Linearity&Power/Pixel(W)&Efficiency (OPS/W)\\
\midrule
APS-P$^2$M\cite{aps_p2m}&Conv, BN, ReLU& 22nm &Analog&Low&Yes &Yes &Low & 0.18$\mu$ &0.4T  \\
% \midrule
 Reconfig\cite{9785835}& Conv& 180nm& Digital& High& No& No& High& 0.025$\mu$-0.11$\mu$ &1.41-3.37T\\
% \midrule
DROIC\cite{zhang202320mum}& Low Light& 180nm& -& -& -& -& -& 0.71$\mu$& -\\
% \midrule
CTIA-CDS\cite{zou2024low}& Low Light& 350nm& -& -& -& -& -& 3.85$\mu$& -\\
% \midrule
  CTIA-IPC (ours) &  Low Light, Conv, BN, ReLU &  22nm&  Digital &  High&   Yes&   Yes&   High&  3.26$\mu$&   3.39G\\
    \bottomrule
\end{tabular}
}
\end{center}
\caption{Comparison of CTIA-IPC with related CTIA and processing-in-pixel works.}
% \caption{Comparison of CTIA-IPC with a pixel array size of 1024 $\times$ 1280, kernel size = 5, stride size = 2, output channel size of 16 with related CTIA and in-pixel computing works.}
\label{tab:comp}
\vspace{-9mm}
\end{table*}

% \begin{table}[t]
% \centering
% \resizebox{\columnwidth}{!}{
% \begin{tabular}{lccccc}
% \toprule
%     \multicolumn{1}{r}{Task}&\multicolumn{1}{c}{Binary seg} &\multicolumn{2}{c}{Parts seg} &\multicolumn{2}{c}{Instrument seg} \\
%     \cmidrule(lr){2-2} \cmidrule(lr){3-4} \cmidrule(lr){5-6}
%     Model &IoU(\%) & IoU(\%) & Dice(\%) & IoU(\%) & Dice(\%)\\
%     \midrule
%     TernausNet \cite{DBLP:journals/corr/abs-1801-05746} & 81.14 & 77.18 & 86.64 & 34.61 & 45.86 \\
%     \midrule
%     Ours & 77.61 & 75.87 & 85.75 & 29.67 & 41.95 \\
%     \bottomrule
% \end{tabular}
% }
% \vspace{0mm} 
% \caption{Performance comparison on three segmentation tasks.} 
% \label{tab:seg}
% \vspace{-9mm} 
% \end{table}

%Circuit Simulation Results: 
\textit{Circuit Simulation Results}: Our modified CTIA pixel circuit is capable of performing MAC operations within a CNN layer while exhibiting improved linearity compared to existing 3T-based in-pixel computing architectures. The SPECTRE simulation results, obtained using 22nm GlobalFoundries FDSOI technology, demonstrate that the MAC operation is a function of the weights stored in the SRAMs and the input photocurrent from the photodiode. Fig. \ref{fig:CTIA_simresults}(a)-(c) represents the analog convolution output voltage on line CBL to show the simulated linearity performance of the MAC operation across a range of shared input photocurrents and SRAM-stored weights. Specifically, as demonstrated in Fig. \ref{fig:CTIA_simresults}(c), the CTIA-based architecture achieves highly linear output responses.  Fig. \ref{fig:CTIA_simresults}(d) illustrates the simulated digital output of the full-range SSADC, clearly demonstrating the high linearity performance of the proposed CTIA-IPC circuit.
The enhanced linearity in our design is primarily attributed to the integration of an OTA within the pixel circuit and the utilization of a capacitive readout-based MAC circuit, which enables accumulation across the CBL. The charge-based computing approach further enhances the signal margin compared to prior current-domain in-pixel processing implementations \cite{aps_p2m,9785835,chen2020pns}. Additionally, we leverage a hardware-algorithm co-design framework to integrate the circuit linearity characteristics into the CNN training process, ensuring that the classification accuracy remains close to state-of-the-art levels for a medical dataset.

Furthermore, we investigate the impact of kernel size on MAC operation linearity. Specifically, Fig. \ref{fig:CTIA_simresults}(e,f,g) confirm that the proposed CTIA-IPC architecture preserves linearity across multiple kernel sizes, demonstrating its flexibility and robustness for biomedical imaging tasks involving diverse convolutional configurations.
% Furthermore, we investigate the impact of kernel size on the linearity of the MAC operation. The simulation results as shown in Fig. \ref{fig:CTIA_simresults}(e,f,g) indicate that when multiple CTIA in pixel cells operate with constant weight and photocurrent: 8 cells for a $3 \times 3$ kernel, 24 cells for a $5 \times 5$ kernel and $48$ cells for a $7 \times 7$ kernel are held constant and 1 cell is swept over different weights and input current — the MAC operation maintains its linearity. This shows our architecture's capability to support multiple kernel sizes efficiently.

A key aspect of our design is the use of a 7-bit global counter with variable step size. The selection of the 4-bit subset, which is fed into the WTC block to match the stored SRAM weights, is dynamically reconfigurable. This would be beneficial to adjust to different lighting conditions for the pixel array as the global counter can select different 4bit subset to feed to the WTC which in turn changes the integration time to $1X$ (bits 3-0 are selected), $2X$ (bits 4-1 are selected) to $8X$ (bits 6-3). This reconfigurable counter step size provides flexibility in adapting to different light conditions while ensuring robust MAC performance.
% For a 3 × 3 kernel, a step size of zero is used, which means that the least significant four bits (3, 2, 1, 0) are selected. In contrast, for 5 × 5 and 7 × 7 kernels, the selection excludes the least significant bit, utilizing bits (4, 3, 2, 1) to maintain a high signal margin. This reconfigurable counter step size provides flexibility in adapting to different kernel sizes while ensuring robust MAC performance.

% when the weight stored in the SRAM was 1111 at a constant photodiode current.
We perform the variability analysis considering both the process and local and global mismatches over 1000 points at a fixed weight and photodiode current.
Fig. \ref{fig:CTIA_simresults}(h) demonstrates the change of voltage in the line CBL. 

%equation:
% \[
% BR = \frac{I}{O} \times \frac{3}{4} \times \frac{12}{N_b}
% \]

% \[
% O = \left( \frac{i - k + 2 \times p}{s} + 1 \right)^2 \times c_o, \quad I = i^2 \times 4
% \]

\textit{Performance Metrics:} To quantify the bandwidth reduction (BR) after the first layer obtained by the CTIA-IPC circuit, let the number of elements in the RGB input image be I ($i^2 \times 4$) and in the output activation map after the pooling layer be O($\left( \frac{i - k + 2 \times p}{s} + 1 \right)^2 \times c_o$). Then, BR can be estimated as: $ \frac{I}{O} \times \frac{3}{4} \times \frac{12}{N_b}\times \frac{1}{p_s^2}$ where $c_o$ denotes the number of output channels of the in-pixel convolutional layer, $p_s$ denotes the pooling stride size and $N_b$ denotes the output bit precision. For a $k =7$ kernel size with $p=0, s=2, c_o=16, N_b=4, p_s=2$  our CTIA-IPC array achieves a BR of 12.08 and it exhibits a $3.26\ \mu$W power consumption per pixel. The compute signal output was sampled from the output of a 6-bit SS-ADC. This work achieves a throughput of 1.98 GOPS with an energy efficiency of 3.39 GOPS/W. Using these performance metrics, we compare our design to prior processing-in-pixel and CTIA-based architectures, as summarized in Table \ref{tab:comp}. Although the reported energy efficiency appears lower compared to existing 3T and 4T-based processing-in-pixel architectures \cite{9785835,aps_p2m}, this result primarily arises from our targeted low light near-tissue application which requires longer exposure time. For biomedical computation BR is the key metric as BR can directly affect the efficiency of the pixel circuit. Our approach employs CTIA pixels optimized specifically for low-light biomedical imaging conditions, inherently limiting throughput and energy efficiency compared to simpler pixel designs; thus, direct efficiency comparisons to earlier processing-in-pixel works are less applicable. However, when benchmarked against existing CTIA pixel architectures \cite{zhang202320mum,zou2024low}, our solution achieves competitive power consumption per pixel while uniquely integrating in-pixel computing capabilities.
% The central contribution of this work is the enhancement of computational linearity within pixel-level operations, significantly improving prediction precision in near-tissue biomedical computing applications.

\begin{table}[t]
\centering
\resizebox{\columnwidth}{!}{ 
\begin{tabular}{lccccc}
\toprule
    \multicolumn{1}{r}{Task}&\multicolumn{1}{c}{Binary Seg} &\multicolumn{2}{c}{Parts Seg} &\multicolumn{2}{c}{Instrument Seg} \\
    \cmidrule(lr){2-2} \cmidrule(lr){3-4} \cmidrule(lr){5-6}
    Model &IoU(\%) & IoU(\%) & Dice(\%) & IoU(\%) & Dice(\%)\\
    \midrule
    TernausNet \cite{DBLP:journals/corr/abs-1801-05746} & 80.14 & 77.18 & 86.64 & 30.10 & 42.34 \\
    % \midrule
    CTIA-IPC 3x (ours) & 78.54 & 75.87 & 85.75 & 29.67 & 41.95 \\
    CTIA-IPC 12x (ours) & 77.59 & 75.91 & 85.81 & 28.58 & 41.03 \\
    \bottomrule
\end{tabular}
}
\vspace{0mm} 
\caption{Performance comparison of the baseline and custom Ternausnet network on the segmentation tasks of the Endovis dataset.} 
\label{tab:seg}
\vspace{-9mm} 
\end{table}

\textit{Task Accuracy}: Building upon our hardware simulation results, we now quantify the accuracy implications of the CTIA-IPC architecture when applied for medical image segmentation tasks on the Endovis dataset. 
We evaluated CTIA-IPC with 3$\times$ and 12$\times$ bandwidth reduction models, where the 3$\times$ model is derived from the 12$\times$ by removing the first-layer max pooling.

As Table \ref{tab:seg} shows, TernausNet achieves 80.14\% IoU in binary segmentation, whereas CTIA-IPC 3$\times$ attains 78.54\% IoU and CTIA-IPC 12$\times$ further reduces this to 77.59\%, corresponding to a modest degradation. 
For parts segmentation, TernausNet reaches 77.18\% IoU and 86.64\% Dice; meanwhile, CTIA-IPC 12$\times$ reaches75.91\% IoU and 85.81\% Dice, retaining approximately 98.3\% and 99.0\% of TernausNet accuracy, respectively. 
% Instrument segmentation shows a similar trend: CTIA-IPC 12× records 28.58\% IoU and 41.03\% Dice versus TernausNet’s 30.10\% IoU and 42.34\% Dice.
Instrument segmentation is more challenging due to its reliance on fine-grained tool articulation features, with TernausNet recording 30.10\% IoU and 42.34\% Dice versus CTIA-IPC 12$\times$’s 28.58\% IoU and 41.03\% Dice.
Note that the $7\times7$ kernel-induced spatial aliasing enhances edge preservation under low-light conditions but introduces minor precision losses in fine-grained tasks like instrument segmentation. Importantly, the $2\times2$ striding in the CTIA convolution module, followed by 4-bit activation quantization and max pooling layer, reduces memory bandwidth by $12\times$. This reduction is critical for real-time processing of 1280×1024 endoscopic video, while the resulting accuracy loss is limited to 1.3\%–2.5\% IoU degradation, with even lower losses observed in the most challenging instrument segmentation task.

% This work achieves a throughput of 1.42 GOPS and an energy efficiency of 3.56 GOPS/W. Using these data we compared our work with previous works that are connected with processing-in-pixel computing and CTIA as shown in Table \ref{tab:comp}. In the table, although the efficiency value may appear lower compared to prior 3T and 4T based processing-in-pixel architectures\cite{9785835,aps_p2m}, this is due to the specific application of our hardware architecture on a robot-surgery-segmentation dataset, which requires a longer computation time. Our circuit employs a CTIA pixel for low light conditions, and the required computation time inherently limits the achievable throughput and energy efficiency, so direct comparisons with previous processing-in-pixel works are less meaningful. However, comparing with existing CTIA pixel works \cite{zhang202320mum,zou2024low}, our work shows good power per pixel while achieving additional in-pixel computing operations. This work specifically targets the application of in-pixel computing in near-tissue computing systems. The main focus of this work was to improve the linearity in in-pixel computing, which improves the precision of predictions when applied to medical datasets.

% \textcolor{red}{Algorithm accuracy results and discussion on hardware-software co-design. (CHENGWEI)}
% \textcolor{red}{Discussion on the comparison table is remaining.}

\section{Conclusion}\label{sec:conclusion}
In this paper, we presented a novel CTIA-IPC architecture that incorporates multi-channel, multi-bit convolution operations alongside batch normalization and quantized ReLU activation, all embedded within the pixel array and the peripheral ADC circuits. Moreover, 3D integration allow our design to store convolutional weights densely within SRAM cells at pixel level without incurring additional area overhead. The algorithm-hardware co-design strategy we adopted, including linearity-aware training and 4-bit quantization, demonstrated strong segmentation performance on the EndoVis dataset, achieving IoU scores close to baseline methods with a subtle accuracy drop ($\sim$1.3\%–2.5\%). Furthermore, our design substantially reduced data bandwidth (12$\times$ reduction), attaining a throughput of 1.98 GOPS and an energy efficiency of 3.39 GOPS/W.

Overall, the CTIA-IPC architecture provides an effective solution for accurate, real-time near-tissue computing, particularly addressing critical medical imaging requirements for high linearity and low-light sensitivity, essential in surgical and diagnostic contexts.

% In this paper, we have purposed a novel in-pixel computing approach leveraging Capacitive Transimpedance Amplifier pixels(CTIA-IPC), which integrates convolution capability within the pixel itself while attaining regular pixel readout capability. Effectively reducing the reliance on external data transfers and off-chip computation. And we have tested our architecture in Endovis Medical dataset achieving accuracy, 12X BR while maintaining an energy efficiency of 3.56 GOPS/W. \textcolor{red}{Needs editing}

% \section*{Acknowledgment}

\bibliographystyle{ieeetr}
\bibliography{sample-base}

\end{document}